\documentclass[aip,rsi,reprint,longbibliography,citeautoscript]{revtex4-2}

\usepackage{amsmath,amssymb} 
\usepackage{bm} 
\usepackage{graphicx} 
\usepackage{comment} 
\usepackage{ulem} 
\usepackage{xspace} 

\usepackage{minted} 
\usepackage{textcomp} 

\usepackage{siunitx} 
\DeclareSIUnit\sample{S}
\DeclareSIUnit\megasample{\mega\sample}
\DeclareSIUnit\kiloamount{k}
\DeclareSIUnit\Siemens{S}
\DeclareSIUnit\mbar{mbar}

\usepackage{enumitem}
\setlist{noitemsep,leftmargin=*,topsep=0pt,parsep=0pt}

\usepackage{xcolor} 
\definecolor{lightgray}{gray}{0.6}
\definecolor{medgray}{gray}{0.4}
\definecolor{mRed}{RGB}{230, 0, 50}
\colorlet{newtextColor}{mRed}

\usepackage{hyperref}
\hypersetup{
colorlinks=true,
urlcolor= blue,
citecolor=blue,
linkcolor= blue,
}

\newif\ifptitle
\newif\ifpnumber
\newcounter{para}
\newcommand\ptitle[1]{\par\refstepcounter{para}
{\ifpnumber{\noindent\textcolor{lightgray}{\textbf{\thepara}}\indent}\fi}
{\ifptitle{\textbf{[{#1}]}}\fi}}

\newif\iftrackchanges

\usepackage{mdframed}
\newmdenv[
  linecolor={\iftrackchanges newtextColor\else white\fi},
  linewidth=2pt,
  topline=false,
  bottomline=false,
  rightline=false,
  skipabove=\topsep,
  skipbelow=\topsep,
  leftmargin=-12pt,
  innertopmargin=0pt,
  innerbottommargin=0pt
]{newtextblock}

\ptitletrue  
\pnumbertrue 

\newcommand{\heng}{School of Engineering \& Applied Sciences, Harvard University, Cambridge, Massachusetts 02138, USA}
\newcommand{\hphys}{Department of Physics, Harvard University, Cambridge, Massachusetts 02138, USA}
\newcommand{\uzhphys}{Department of Physics, University of Zurich, Winterthurerstrasse 190, CH-8057 Zurich, Switzerland}
\newcommand{\instvalencia}{Instituto de Ciencia Molecular, Universitat de València, 46980 Paterna, Spain}
\newcommand{\ugeneva}{Department of Quantum Matter Physics, University of Geneva, 24 Quai Ernest-Ansermet, CH-1211 Geneva, Switzerland}

\newcommand{\mytitle}{Reversible nanoscale patterning of WTe$_2$ with a scanning tunneling microscope}

\newcommand{\wte}{WTe$_2$\xspace}

\newcommand{\basPatt}{base pattern\xspace}
\newcommand{\basPatts}{base patterns\xspace}

\begin{document}

\title{\mytitle}

\author{Kevin Hauser}
\affiliation{\hphys}
\affiliation{\uzhphys}

\author{Danyang Liu}
\affiliation{\uzhphys}

\author{Berk Zengin}
\affiliation{\uzhphys}

\author{Jens Oppliger}
\affiliation{\uzhphys}

\author{Samuel Mañas-Valero}
\affiliation{\instvalencia}

\author{Catherine Witteveen}
\affiliation{\uzhphys}
\affiliation{\ugeneva}


\author{Fabian O. von Rohr}
\affiliation{\ugeneva}

\author{Jennifer E. Hoffman} 
\affiliation{\hphys}
\affiliation{\heng}

\author{Fabian D. Natterer}
\email[]{fabian.natterer@physik.uzh.ch}
\affiliation{\uzhphys}

\date{\today}

\begin{abstract}
Manipulating the lattice structure of ferroelectric quantum materials enables their use in low-power electronic devices, including field-effect transistors.
\wte is a Weyl-semimetal candidate and ferroelectric, both properties arising from the reduced crystal symmetry of its T$_\mathrm{d}$ ground state. The T$_\mathrm{d}$ crystal phase results from a Peierls distortion of the 1T parent structure and an interlayer shift.
While experiments in \wte have established ferroelectric switching and transient control of the predicted topological phase via ultrafast excitations, persistent electronic changes on the nanometer scale remain elusive.
Here, we demonstrate that current pulses applied via scanning tunneling microscopy can both write and erase persistent nanometer-scale patterns on the surface of \wte. 
These patterns consist of apparent picometer in-plane and out-of-plane atomic displacements, accompanied by changes to the local density of states. 
The out-of-plane displacements further modulate the Peierls-like distortion present in \wte, while the in-plane displacements are indicative of ferroelectric switching.
The induced patterns can be repositioned and erased, suggesting a nanoscale handle on the ferroelectric properties of \wte. 

\end{abstract}

\maketitle

\section{\label{sec:Intro}Introduction}

Power-efficient field-effect transistors have been proposed based on functionalized ferroelectric (FE)\cite{ZhangNatRevMater2023} or Weyl semimetal (WSM) phases\cite{RocchinoNatCommun2024}. 
However, such next-generation, high-density devices will require persistent, reversible control over these electronic phases at the nanometer scale.
Tungsten ditelluride (\wte) provides a uniquely versatile platform for such control because its orthorhombic lattice naturally breaks inversion symmetry, thereby enabling both its polar and topological properties. \wte is a rare ferroelectric metal\cite{FeiNat2018, YangJPCL2018, LiuNano2019}, a WSM candidate in its bulk form\cite{SoluyanovNat2015, BrunoPRB2016, LinACSNano2017, LvPRL2017, LiNatCommun2017}, and a quantum spin Hall insulator in the monolayer limit\cite{QianSci2014, TangNatPhys2017}.
Transport and piezo force microscopy have shown micron-scale control over its ferroelectric order\cite{FeiNat2018, SharmaSciAdv2019}, and pump-probe experiments have shown transient structural tunability of the predicted WSM phase on the picosecond time scale\cite{SieNat2019, SoranzioPRR2019, SoranzioNPJ2D2022}. Yet control over persistent, nanometer-scale structural changes and the resulting electronic phases remains elusive.

\begin{figure*}
    \includegraphics[width=\textwidth]{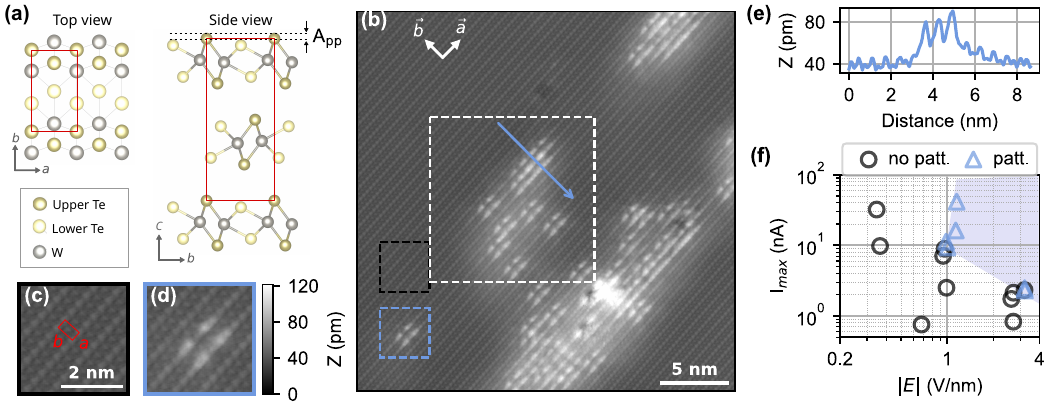}
 	\caption{
    Basic appearance of pattern. 
    \textbf{(a)} Top and side view of \wte crystal structure. Unit cell is shown in red. We depict Te atoms at different heights in shades of yellow. 
    The top view shows the W and Te atoms of the surface \wte layer.
    $A_\mathrm{pp}$ illustrates the height difference between the upper and lower Te atoms at the surface. 
    \textbf{(b)} Large topography showing both an unperturbed area and regions with patterns of various sizes. Recorded at a sample bias $V_\mathrm{s}\!=\!\qty{300}{\milli\volt}$ and set current $I_\mathrm{s}\!=\!\qty{500}{\pA}$. The white dashed box indicates the region under analysis in Fig.~\ref{fig:distort}. 
    \textbf{(c)} and \textbf{(d)} show zoom-ins of the pristine \wte surface and the smallest unit of the created pattern, respectively. Locations are indicated in (b) by the black and blue dashed rectangles. The unit cell is shown in red.
    \textbf{(e)} Height trace along the blue arrow in (b) illustrates the change in corrugation across larger patches of coalescing \basPatts.
    \textbf{(f)} Overview of successful and unsuccessful patterning events dependent on the maximal tunneling current $I_\mathrm{max}$ and electric field $|\vec{E}|$ applied between STM tip and sample. Blue shaded region serves as guide to the eye where patterning is possible. 
    }
	\label{fig:topo}
\end{figure*}

\wte is a layered van der Waals material whose non-centrosymmetric T$_\mathrm{d}$-crystal phase (Fig.~\ref{fig:topo}(a)) combines a Peierls-like intralayer distortion of the ideal 1T lattice with an orthorhombic interlayer stacking\cite{BrownActaCryst1966, AugustinPRB2000, AliNature2014, QianSci2014, LeeSciRep2015}.
As a result, Te atoms within a chalcogen layer assume different heights ($A_\mathrm{pp}$ in Fig.~\ref{fig:topo}(a)). This naturally occurring distortion leads to quasi-1D W--W chains along the $\vec{a}$ axis that act as transport channels\cite{AugustinPRB2000, JhaAIPA2018}. 
FE and the predicted WSM phase result from the non-centrosymmetric crystal structure and both can be altered through in-plane shifts of subsequent van der Waals layers\cite{YangJPCL2018, SoluyanovNat2015, SieNat2019}.
It is reported that the interlayer stacking depends on the charge concentration around the outer Te atoms\cite{KimPRB2017}. Consistent with this, horizontal sliding of individual layers has so far been achieved by exciting carriers through optical pumping\cite{SieNat2019} or by applying a large out-of-plane electric field\cite{FeiNat2018}.

Here we report the persistent and reversible creation of nanoscale patterns at the surface of \wte induced by current pulses from a scanning tunneling microscope (STM). 
We show that these patterns consist of apparent in- and out-of-plane atomic displacements in the form of a lateral movement for the first and changes to the Te row height contrast for the latter.
Using spatial lock-in analysis\cite{LawlerNature2010, HytchUltramicroscopy1998}, we further quantify the apparent in-plane and out-of-plane atomic displacements from our STM topographies.
The apparent displacements along the crystallographic $\vec{b}$ axis, we observe, are comparable to interlayer shifts associated with ferroelectric switching\cite{YangJPCL2018} and changes to the predicted WSM phase\cite{SieNat2019}. 
We demonstrate the ability to reposition and erase the induced patterns, illustrating control on the nanometer scale.

\section{Methods}

The study reported here was conducted on \wte crystals synthesized by two methods: Cl$_2$‑assisted chemical vapor transport and the Te-flux method (Appendix~\ref{sec:app:cryst-grow}).
Bulk \wte crystals were cleaved in situ at room temperature and a base pressure of $\qty{5e-10}{\mbar}$. All measurements were performed on a low-temperature, bath-cryostat STM from Createc under ultra-high vacuum. The STM temperature was \qty{4.2}{\kelvin} if not indicated otherwise. Etched W tips were annealed with an electron beam to remove the oxide layer and characterized on an Au$(111)$ surface to verify sharpness and confirm the featureless density of states. 

\section{Results}

\subsection{Create pattern}

The unperturbed \wte surface (upper-left portion of Fig.~\ref{fig:topo}(b) and zoom-in (c)) agrees with previous STM studies\cite{ZhengPRL2016, LinACSNano2017, LiPRB2016, YuanPRB2018, ZhangPRB2017, KawaharaAPEX2017, SanchesPRB2025}. Our topographies resolve the upper and lower Te atoms of the top chalcogen layer\cite{ZhengPRL2016, KawaharaAPEX2017}. 
Fig.~\ref{fig:topo}(d) illustrates a single unit of the induced pattern (\basPatt), resembling four-tire impressions. 
Its dominant feature is a protrusion of the upper Te atoms, which spans a $6\,\vec{a}\!\times\!\vec{b}$ area.
Fig.~\ref{fig:topo}(b) shows how multiple \basPatts are reproduced across the surface and how they coalesce to form larger patches. These patches show two distinct height signatures in the cross section in Fig.~\ref{fig:topo}(e). 
First, the corrugation $A_\mathrm{pp}$ between the upper and lower Te atoms is increased and appears as sharp peaks that follow the $\vec{b}$-axis periodicity.
Second, there are fainter elevations spanning a few nanometers and encompassing the sharp features. 
The height of the coalescing patterns is more than just the sum of multiple \basPatts, suggesting a non-linear overlap.
We further note that the larger patches appear to run preferentially along the Te rows ($\vec{a}$ axis), as seen in Fig.~\ref{fig:topo}(b).

To better understand the patterning process, we varied the bias voltage $V_\mathrm{s}$ applied between the tip and sample, and the maximal tunneling current $I_\mathrm{max}$.
We observe that the patterns arise from the combined effect of high currents and elevated electric fields, as summarized in Fig.~\ref{fig:topo}(f). 
An electric-field threshold of $|\vec{E}|\!\sim\!\qty{1}{\volt\per\nm}$ must be exceeded for pattern creation. Above this value, stronger fields lower the required current. 
The formation process has a stochastic component, as overlapping symbols for successful and failed attempts demonstrate. 
Higher applied currents increase the lateral size of the resulting patterns (not shown).
We did not observe a dependence on bias polarity. 
Details on how Fig.~\ref{fig:topo}(f) is compiled can be found in Appendix~\ref{sec:app:patt}.
It is noteworthy that $I_\mathrm{max}$ and $V_\mathrm{s}$ employed for patterning are outside of the range typically used for STM/STS measurements on \wte (see Appendix~\ref{sec:app:ivmap}).

\begin{figure*}
    \includegraphics[width=\textwidth]{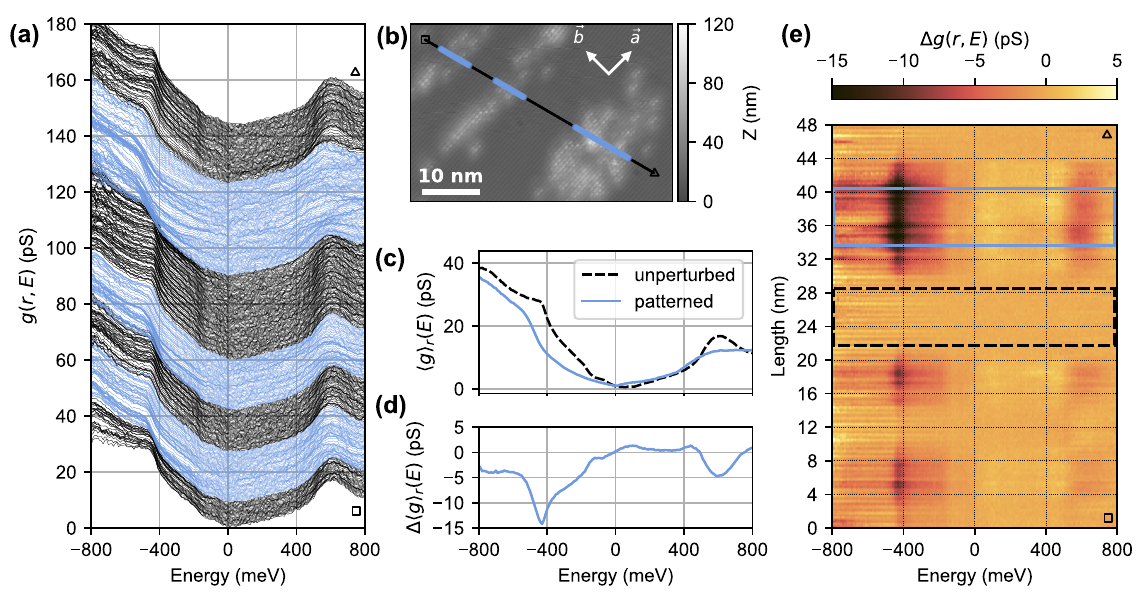}
 	\caption{
    Local density of states of the patterned \wte surface.
    \textbf{(a)} Waterfall plot of the measured differential conductance $g(r,E)$. The blue shaded spectra indicate the location of the patterns determined by a depletion in LDOS at \qty{-400}{\meV} (Details in Appendix~\ref{sec:app:pattLDOS}). 
    The directionality is indicated  by the black triangle and square on the right side, as well as in (b) and (e). 
    \textbf{(b)} Location of spectra shown in (a). Blue segments correspond to the ones in (a). Topography was recorded with $V_\mathrm{s}\!=\!\qty{300}{\milli\volt}$ and $I_\mathrm{s}\!=\!\qty{2}{\nano\ampere}$.
    \textbf{(c)} Spatially averaged conductance $\langle g\rangle_r(E)$ of the patterned and unperturbed regions highlighted by the dashed boxes in (e).
    \textbf{(d)} Difference in conductance between LDOS spectra in (c). 
    \textbf{(e)} Conductance difference at every point to averaged unperturbed LDOS spectrum $\langle g \rangle_{r,0}(E)$ (black curve in (c).), obtained through $\Delta g(r,E)\!=\!g(r,E)-\langle g \rangle_{r,0}(E)$.
    LDOS Spectra in (a) were acquired with a lock-in frequency $f\!=\!\qty{887}{\hertz}$ and modulation $V_\mathrm{pp}\!=\!\qty{2}{\milli\volt}$ at $V_\mathrm{s}\!=\!\qty{1}{\volt}$ and $I_\mathrm{s}\!=\!\qty{2}{\nano\ampere}$ and smoothed with a Gaussian filter with $\sigma\!=\!\qty{1}{\meV}$ in post-processing. The waterfall plot uses an offset of \qty{0.3}{\pico\Siemens} between spectra. 
    }
	\label{fig:el}
\end{figure*}

\subsection{Electronic structure}

To better understand the electronic structure changes induced by the patterns, we used scanning tunneling spectroscopy (STS). We measure the local density of states (LDOS) in the form of the differential conductance $g(r,E)\!\equiv\!dI/dV$ across a line traversing multiple patterned regions (Fig.~\ref{fig:el}(a) and (b)).
Averaged LDOS spectra are shown in Fig.~\ref{fig:el}(c).
The unperturbed surface exhibits a pronounced edge at \qty{-450}{\meV}, a broad peak near \qty{550}{\meV}, and multiple kinks between \qty{-100}{\meV} and \qty{150}{\meV} corresponding to the conduction and valence band extrema. These LDOS features are in agreement with previous studies\cite{LiPRB2016, ZhangPRB2017}. 
The predicted Weyl points are located around \qty{55}{\meV} \cite{SoluyanovNat2015, LinACSNano2017}.
The patterned region modifies the LDOS in all of these energy ranges, as illustrated in Fig.~\ref{fig:el}(d), where we plot the conductance difference $\Delta \langle g(E)\rangle_r$ between the two spatially averaged LDOS spectra shown in Fig.~\ref{fig:el}(c).
The strongest LDOS depletion occurs near \qty{-450}{\meV}, while spectral weight is enhanced near the band extrema and the predicted Weyl points. 
In Fig.~\ref{fig:el}(e), we plot the conductance difference $\Delta g(r,E)$ between every spectrum $g(r,E)$ and the averaged LDOS of the unperturbed surface $\langle g \rangle_{r,0}(E)$. LDOS depletions appear dark red, enhancements yellow.  All three patterned segments exhibit LDOS modifications, which become more pronounced as the regions become wider.

Height information measured with STM combines topographic and electronic contrast.
By correlating height information with spatially resolved LDOS spectra, we distinguish structural from electronic contributions.  
As shown in Fig.~\ref{fig:topo}(e), two distinct height signatures are evident: (i) enhanced atomic corrugation of the Te rows, and (ii) a fainter height increase spanning a nm-range around the patterned regions. 
The latter coincides with enhanced LDOS between \qty{0}{\meV} and \qty{300}{\meV}, as visible in Fig.~\ref{fig:el}(e), implying an electronic rather than structural origin. 
In contrast, the short-range corrugation lacks a corresponding LDOS modulation, indicating that it arises from atomic out-of-plane displacements. 
For comparison, LDOS variations on the unit-cell length scale can in principle occur, but are only pronounced below \qty{-400}{\meV}.

\begin{figure*}
    \includegraphics[width=\textwidth]{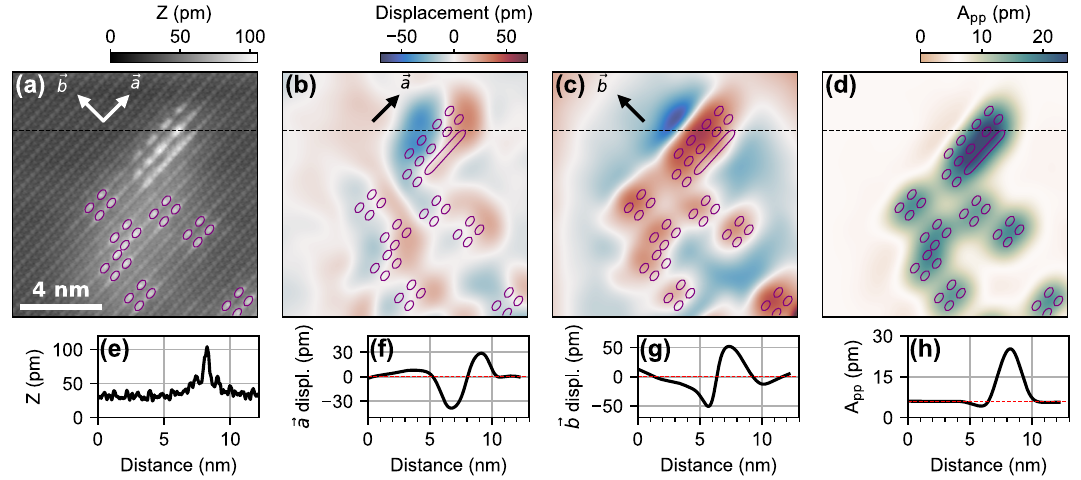}
    \caption{
        Spatial lock-in analysis to resolve apparent lattice distortions.
        \textbf{(a)} Topography 
        corresponding to white dashed rectangle in Fig.~\ref{fig:topo}(c). The four tire tracks of the \basPatts are marked by purple circles to provide a spatial reference between panels (a) to (d). 
        Lattice directions $\vec{a}$ and $\vec{b}$ are indicated in the top left corner. 
        \textbf{(b)} and \textbf{(c)} show apparent in-plane displacements along $\vec{a}$ and $\vec{b}$, extracted via spatial lock-in analysis. 
        The plotted displacement values are obtained from the spatial lock-in phase map by rescaling with the unit cell length of the respective lattice direction.
        Arrows indicate displacement direction.
        \textbf{(d)} Apparent variation in Te-height corrugation $A_\mathrm{pp}$, defined as the spatial lock-in amplitude of the first-order $\vec{b}$ Bragg peak. 
        \textbf{(e)} to \textbf{(h)} show line cuts along the horizontal dashed lines in panels (a) to (d) The red dashed lines in (f) to (h) indicate the unperturbed state.
        }
	\label{fig:distort}
\end{figure*}

\subsection{Spatial lock-in analysis}

We performed spatial lock-in analysis\cite{LawlerNature2010, HytchUltramicroscopy1998} on the topography in Fig.~\ref{fig:distort}(a).
To do this, we multiplied the image by sinusoidal reference functions matching the periodicities of the first-order Bragg peaks along $\vec{a}$ and $\vec{b}$, respectively. We low-pass filtered the result with a Gaussian ($\sigma_x\!=\!\sigma_y\!=\!\qty{360}{\pm}$).  
This analysis yielded spatially varying phases and amplitudes of the selected Fourier components, which we then used to construct the displacement maps shown in Fig.~\ref{fig:distort}(b)–(d).

The $\vec{a}$-axis displacement map (Fig.~\ref{fig:distort}(b) and (f)) shows positive and negative domains separated by a boundary that runs through the center of each \basPatt.
The maximum displacement within the field of view shown here is $\sim\!\pm\qty{40}{\pm}$ along $\vec{a}$ (12\,\% of the unit cell).  
By contrast, the $\vec{b}$-axis map (Fig.~\ref{fig:distort}(c) and (g)) shows a unidirectional shift at the top and an opposing displacement around the patterns. The larger region with coalescing patterns shows a displacement magnitude of up to $\sim\!\pm\qty{50}{\pm}$ along $\vec{b}$ (8\,\% of the unit cell).

The amplitude map in Fig.~\ref{fig:distort}(d) shows the spatially varying corrugation $A_\mathrm{pp}$ of the upper and lower Te rows making up the top chalcogen layer.  
On top of the patterns, the corrugation rises to \qty{25}{\pm}, roughly four times the value for the unperturbed surface, with the latter indicated by the red dashed line in Fig.~\ref{fig:distort}(h).  
Furthermore, a narrow collar of reduced corrugation ($A_\mathrm{pp}\!\sim\!\qty{3}{\pm}$) surrounds the patterns, indicating that corrugation variations extend $1\!-\!\qty{2}{\nm}$ beyond the sharp topographic protrusions.


\subsection{Manipulation}

\begin{figure*}
    \includegraphics[width=\textwidth]{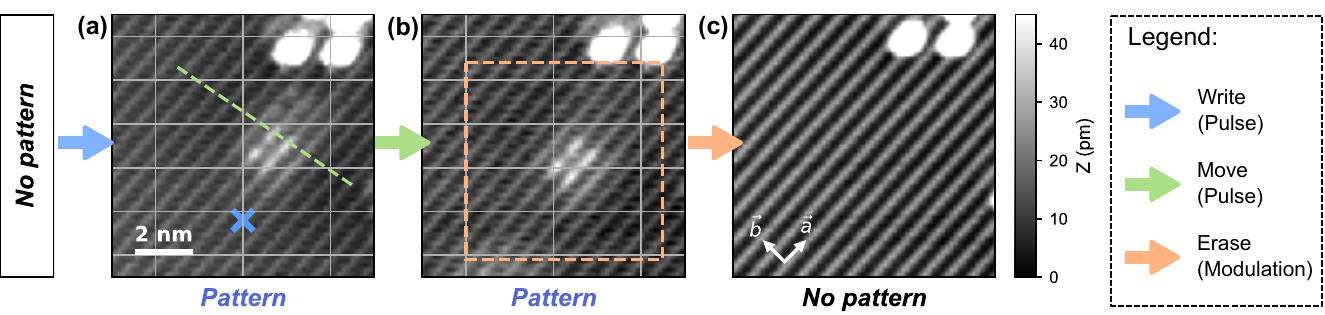}
 	\caption{
    Base pattern movement and deletion.
    \textbf{(a)} to \textbf{(c)} show constant-current topographies recorded after write (a), move (b), and erase (c) operations (green, blue, and orange arrows explained in the legend on the right). 
    The blue cross and green dashed line in (a) indicate the location of the  spectra used for pattern creation and movement. The orange dashed rectangle in (b) indicates the location of the erase operation.
    Grid lines are included in (a) and (b) to illustrate the movement of the pattern along the Te row.
    The bright white dots on the top right are surface contaminants that act as positional reference between topographies. Lattice directions $\vec{a}$ and $\vec{b}$ are indicated in panel (c). 
    ($V_\mathrm{s}\!=\!\qty{500}{\milli\volt}$, $I_\mathrm{s}\!=\!\qty{700}{\pA}$).
    }
	\label{fig:fig4}
\end{figure*}

In Fig.~\ref{fig:fig4}, we show an operation sequence in which we created, translated, and then erased a \basPatt.  
We achieved translation along the $\vec{a}$ direction by applying current pulses along the green dashed line in Fig.~\ref{fig:fig4}(a). $I_\mathrm{max}$ and $V_\mathrm{s}$ were chosen close to the boundary of the blue shaded region in Fig.~\ref{fig:app:pattmapIV}. 
Because pattern writing is independent of bias polarity, current pulses cannot be used for deletion. 
Instead, we applied a bias modulation with $f\!=\!\qty{1}{\kHz}$ and $V_\mathrm{pp}\!=\!\qty{750}{\mV}$ at a bias offset $V_\mathrm{off}\!=\!\qty{250}{\mV}$ to the tip junction at constant height, producing an oscillating current. 
With the bias modulation applied, we gradually moved the tip closer to the surface in discrete steps until $I_\mathrm{max}\!=\!\qty{6}{\nA}$. 
We repeated this process across a grid of points spanning the area indicated by the orange dashed box in Fig.~\ref{fig:fig4}(b) to make the pattern disappear (Fig.~\ref{fig:fig4}(c)). 
We assume that the oscillating electric field and current smooths out the lattice displacement associated with the pattern as we move across the grid, thereby restoring the unperturbed surface.

\section{Discussion}

Optical pump-probe experiments show that in \wte, excited carriers stimulate lattice vibrations through displacive excitation of coherent phonons (DECP)~\cite{DaiPRB2015, DruekePRB2021}. 
Electrons and holes induced by STM tunneling also act as excited carriers.
In the regime of large tunneling currents as outlined in Fig.~\ref{fig:topo}(f), STM yields excited-carrier concentrations comparable to those in DECP pump-probe experiments due to the atomic-scale tunneling area and the quasi-1D transport properties (see Appendix~\ref{sec:app:excite} for an order-of-magnitude estimation).
Together with surface electric fields on the order of $\qty{1}{\volt\per\nm}$, these conditions may stabilize the atomic displacements reported here.  
This interpretation is consistent with both the pattern aligning preferentially along the low-resistance $\vec{a}$ axis and the electric-field threshold observed in Fig.~\ref{fig:topo}(f).
Because \wte is a van der Waals material, local excitations caused by STM tunneling currents are likely confined to the top layer; the resulting dynamical asymmetry relative to the subsurface layers may further stabilize distortions.

Since the in-plane atomic displacement is not expected to propagate into deeper layers, the shifts in Fig.~\ref{fig:distort}(b) and (c) can be interpreted as a displacement of the top layer relative to the bulk.  
Along $\vec{b}$, the displacement changes from $-\qty{50}{\pm}$ to $+\qty{50}{\pm}$ over a distance of $\qty{2}{\nm}$ (Fig.~\ref{fig:distort}(g)).
This magnitude is comparable to the predicted interlayer sliding of $\qty{72}{\pm}$ required for ferroelectric reversal~\cite{YangJPCL2018}, suggesting a local change in polarization across the pattern. 
Unlike the predicted switching, which occurs strictly along $\vec{b}$~\cite{YangJPCL2018}, the displacement observed here includes an $\vec{a}$ component, implying a more complicated reorientation of the polarization vector.

A displacement of $\qty{2.2}{\pm}$ along $\vec{b}$ has been reported to drive a topological phase transition by annihilating the suspected Weyl points\cite{SieNat2019}.  
Fig.~\ref{fig:distort}(b) reveals displacements exceeding this threshold, implying that the predicted WSM phase could be locally suppressed.  
Furthermore, our STS measurements (Fig.~\ref{fig:el}(f)) show modifications to the LDOS induced by the pattern around $E\!=\!\qty{+55}{\meV}$, the energy of the suspected Weyl points and surface Fermi arcs~\cite{SoluyanovNat2015, LinACSNano2017}.  
However, the available energy resolution is insufficient to determine these LDOS changes more precisely.

\section{Conclusion}

We demonstrate the controlled creation and deletion of apparent lattice displacements on the surface of \wte, using current pulses from an STM tip. 
The induced patterns, spanning a few unit cells up to tens of nanometers, consist of apparent in-plane atomic shifts and variations to the Te atom corrugation. 
The in-plane displacements are of comparable magnitude as ferroelectric switching and reported alterations to the predicted Weyl semimetal phase in \wte. The induced change to the Te height corrugation suggests a local variation in the Peierls-like distortion naturally present in \wte. Using STS, we further show that the pattern locally alters the electronic structure. 
Our findings of nanoscale manipulation of the lattice and electronic structure in \wte may support future studies exploring the engineering of ferroelectric or topological order.

Future investigations could utilize Kelvin probe force spectroscopy\cite{AlbrechtPRB2015} to resolve the local variations in contact potential difference to confirm the suspected ferroelectric switching across the pattern.
To better understand the changes to the LDOS induced by the pattern, quasiparticle interference (QPI) mapping~\cite{hoffmanScience2002, mcelroyNature2003} could be performed, which provides nanometer-scale spatial resolution and enables direct correlation between structural features and $k$-space modifications.  
QPI has already been employed to investigate the predicted WSM phase in \wte~\cite{LinACSNano2017, ZhangPRB2017, YuanPRB2018, SanchesPRB2025}, and applying it here could reveal local momentum-space changes around the patterned regions.

\section*{Acknowledgments}
We acknowledge support from the Swiss National Science Foundation under Grant Nos. 200021-232187, 206021-213238, PP00P2-211014, and the Air Force Office of Scientific Research (AFOSR) Multidisciplinary University Research Initiative (MURI) grant FA9550-21-1-0429.
We thank Raagya Arora, Daniel Larson, Danny Bennet, and Gal Tuvia for fruitful discussions.

\section*{Data Availability}
The data that support the findings of this study are available from the corresponding author upon reasonable request.

\newcounter{appendix_count}
\setcounter{appendix_count}{1}

\newcommand{\appendixsection}[2]{
    \section{#1} \label{#2}
    \stepcounter{appendix_count}
}

\appendix

\appendixsection{Crystal growth parameters}{sec:app:cryst-grow}

Two batches of \wte single crystals were used for this study. The results presented here were measured on crystals grown by chemical vapor transport (CVT) with chlorine as the transport agent. 
Control experiments were then conducted on \wte crystals grown by the Te-flux method.

\ptitle{CVT batch}
For the CVT growth, polycrystalline \wte was first prepared by reacting stoichiometric W (99.99\%) and Te powder (99.997\%) sealed in an evacuated quartz ampoule (pressure $\sim\qty{5e-5}{\mbar}$; length \qty{25}{\centi\metre}; inner diameter \qty{1.5}{\centi\metre}). 
The ampoule was heated from room temperature to \qty{910}{\celsius} at \qty{5}{\celsius\per\minute}, held for 9\,days, and allowed to cool to room temperature. Single crystals were then obtained by sealing \qty{1}{\gram} of this polycrystalline \wte together with Cl$_2$ transport agent (\qty{1.34}{\milli\mole}) in a second evacuated quartz ampoule (pressure $\sim\qty{1.3e-5}{\mbar}$; length \qty{50}{\centi\metre}; inner diameter \qty{1.5}{\centi\metre}) placed in a three-zone furnace with the material at one end. 
Two zones were heated to \qty{650}{\celsius} at \qty{0.5}{\celsius\per\minute}; the source zone was then raised to \qty{750}{\celsius} over \qty{12}{\hour} to establish a \qty{100}{\celsius} gradient. 
The temperature was held for 7\,days before being cooled down naturally. Energy-dispersive X-ray spectroscopy measured W = $40.9\pm1.8$\,\% and Te = $59\pm2$\,\%. The powder X-ray diffraction pattern was refined using ICSD~14348 with lattice parameters $a=6.277(6)$\,\AA, $b=3.497(3)$\,\AA, and $c=14.07(1)$\,\AA. The obtained results were in accordance with those reported in the literature\cite{BrownActaCryst1966}.

\ptitle{Te-flux batch}
For the Te-flux growth, tungsten ingots (99.95\%) and tellurium lumps (99.999+\%) were combined in a 1:20 molar ratio and sealed under vacuum in a quartz tube. 
The reagents were heated to \qty{1000}{\celsius} in \qty{10}{\hour}, held for \qty{24}{\hour}, and cooled to \qty{600}{\celsius} in \qty{100}{\hour}. 
At \qty{600}{\celsius} the excess Te flux was removed by centrifugation. The ampoule was air-quenched, and the crystals were annealed at \qty{400}{\celsius} for \qty{12}{\hour} to remove residual Te.

\appendixsection{Compile parameter map in Fig.~\ref{fig:topo}(f)}{sec:app:patt}

\ptitle{Patterning operation} 
We used the STM's bias spectroscopy mode, typically used to measure current-voltage (I--V) traces, to perform the patterning operation. 
Although bias pulses with disabled height-feedback would produce the same effect, spectroscopy was preferred because it records the tunneling current and sample bias values for post-analysis. 

\begin{figure}[t] 
  \centering
  \includegraphics[width=0.8\columnwidth]{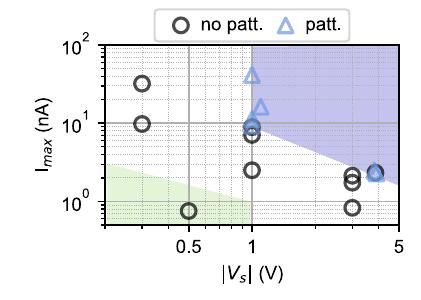}
  \caption{Patterning-parameter map with the tip-sample bias $V_s$ shown on the $x$-axis instead of the calculated electric field $E$ (Fig.~\ref{fig:topo}(f)). Blue-shaded region serves as guide to the eye where patterning is possible. The green-shaded region highlights STM setpoint values used in previous studies\cite{YuanPRB2018, ZhengPRL2016, LinACSNano2017, PengNatCommun2017}.
  }
  \label{fig:app:pattmapIV}
\end{figure}

\ptitle{Compiling patterning parameter map} 
The patterning-parameter map shown in Fig.~\ref{fig:topo}(f) was compiled as follows: $I_\mathrm{max}$ represents the maximal tunneling current present during the patterning operations---irrespective of bias polarity. This value was read directly from the measured I--V spectra. 
For comparison, a patterning map showing the measured maximal current $I_\mathrm{max}$ and the corresponding tip-sample bias $V_\mathrm{s}$ is shown in Fig.~\ref{fig:app:pattmapIV}.
The electric field on the $x$-axis in Fig.~\ref{fig:topo}(f) was estimated using 

\begin{equation}
    |\vec{E}| = \frac{V_\mathrm{s}\!+\!V_\mathrm{CPD}}{z} \approx \frac{V_s}{z}, 
\end{equation}

\noindent with the applied sample bias $V_\mathrm{s}$, the contact potential difference $V_\mathrm{CPD}$, and the tip–sample distance $z$. 
We neglected $V_\mathrm{CPD}$ since it is much smaller than the applied bias ($V_\mathrm{CPD}\!\approx\!\qty{0.1}{\volt}$, assuming tip and sample work functions of $\phi_\mathrm{tip}\!=\!\qty{4.5}{\eV}$ (polycrystalline W~\cite{MichaelsonJAP1977}) and $\phi_\textrm{\wte}\!=\!\qty{4.6}{\eV}$~\cite{KwonApplSurfSci2020}).
Note, the comparable work functions are consistent with the observation that the pattern creation is bias-polarity independent. 
STM provides precise control of relative height changes in the picometer range but does not provide a direct measure of the absolute tip–sample distance $z$.  
Therefore, we used 

\begin{equation}
    z = z_0 + \Delta z
\end{equation}

\noindent by assuming a tip–sample height of $z_0\!=\!\qty{500}{\pico\meter}$ at a setpoint of $V_\mathrm{0}\!=\!\qty{500}{\mV}$ and $I_\mathrm{0}\!=\!\qty{500}{\pA}$, and calculating the relative height difference $\Delta z$, for the different tunneling setpoints of the various patterning trials in Fig.~\ref{fig:topo}(f), as 

\begin{equation}
    \Delta z(I)|_{V_0} = \frac{\log(I/I_0)}{2\kappa},
\end{equation}

\noindent with the decay length $\kappa\!=\!\qty{11}{\per\nm}$, calculated from 

\begin{equation}
    \kappa\!=\!\sqrt{2m_e \bar{\phi}} / \hbar
\end{equation}

\noindent where $\bar{\phi}$ is the average work function between tip and sample.

\ptitle{Factors of uncertainties for estimation}
Because $z_0$ is not known precisely, the absolute electric-field values carry a considerable uncertainty. 
However, because $\Delta z$ together with $V$ sets the horizontal coordinate of each point in Fig.~\ref{fig:topo}(f), the relative positions of the points with respect to each other remain reliable.
Another limitation of our electric-field estimation is that pattern creation also depends on the local field near the STM apex during tunneling.  
Because the tip shape beyond the atoms contributing to the tunneling process is generally unknown, the exact field distribution is not known either.  
Consequently, the reported absolute electric-field values should be regarded as estimates.

\appendixsection{Patterning parameter vs. typical STM setpoint values in \wte}{sec:app:ivmap}
To provide an experimentally accessible parameter map for patterning, we generated Fig.~\ref{fig:app:pattmapIV}, which shows the applied sample bias $V_\mathrm{s}$ on the horizontal axis rather than the electric field $|\vec{E}|$ shown in Fig.~\ref{fig:topo}(f). 
Comparing with typical setpoint values $I_\mathrm{s}$ and $V_\mathrm{s}$ used in previous STM studies (green-shaded area)\cite{YuanPRB2018, ZhengPRL2016, LinACSNano2017, PengNatCommun2017} shows that these values do not overlap with the region in which we observe patterning.
This suggests that the $I_\mathrm{s}$--$V_\mathrm{s}$ parameter regime in which patterning occurs may not have been explored in previous STM studies.


\appendixsection{Quantify spatial extent of pattern from LDOS change}{sec:app:pattLDOS}
To quantify the spatial extent of the patterned regions, we define a conductance threshold that separates the line in Fig.~\ref{fig:el}(b) into patterned and unperturbed regions.  
We use the LDOS at $E\!=\!\qty{400}{\meV}$ since it shows the most pronounced spectral-weight change as can be seen in Fig.~\ref{fig:el}(e). 
An unperturbed region is identified manually. Across the selected, unperturbed region, we estimate an average conductance value at $E\!=\!\qty{400}{\meV}$ and its standard deviation $\sigma$.
The patterned region is then defined as any point along the recorded line LDOS that deviates more than $5\sigma$ from the average value.
For visual reference, the segments, which deviate, are marked in blue in Fig.~\ref{fig:el}(a) and (b).

\appendixsection{Estimate excited carrier density from tunneling current}{sec:app:excite}

Here, we present an order-of-magnitude estimation of the resulting excited carrier density $n$ created by our tunneling current.
We assume that the tunneled electrons or holes enter \wte as excited carriers, which then dissipate along the W--W\ chains. To estimate $n$, we use a steady-state rate equation,

\begin{equation}
    n = \frac{J \tau}{e d}
    \label{eq:currDensity}
\end{equation}

where $J$ is the current density, $\tau$ is the excited carrier lifetime, $d$ is the approximated distance across which the excited carriers relax, and $e$ is the elementary charge. 
We assume $\tau\!=\!\qty{10}{\pico\second}$ according to reported electron-hole recombination times\cite{DaiPRB2015} and $d\!=\!\qty{10}{\nm}$ estimated as the length scale associated with our pattern (see Fig.~\ref{fig:el}(b) for example). 
We obtain a current density of $J=I/A_\mathrm{eff}=\qty{5e6}{\ampere\per\square\centi\metre}$ for a tunneling current of $\qty{10}{\nano\ampere}$ and an effective area of $A_\mathrm{eff}=\vec{b}/2 \times \vec{c}/2 = \qty{0.2}{\square\nm}$.
Using Eq.~\ref{eq:currDensity} we obtain an excited carrier concentration of $n\!=\!\qty{2.5e20}{\per\cubic\centi\metre}$, which is larger than the reported value of $n\!=\!\qty{3e18}{\per\cubic\centi\metre}$ for an onset of coherent phonon excitation\cite{DaiPRB2015}. 



\section*{\label{sec:Refs}References}

\bibliography{refs}

@article{AliNature2014,
	author = {Ali, Mazhar N. and Xiong, Jun and Flynn, Steven and Tao, Jing and Gibson, Quinn D. and Schoop, Leslie M. and Liang, Tian and Haldolaarachchige, Neel and Hirschberger, Max and Ong, N. P. and Cava, R. J.},
	title = {Large, non-saturating magnetoresistance in {WTe2}},
	journal = {Nature},
	year = {2014},
	volume = {514},
	number = {7521},
	pages = {205--208},
	doi = {10.1038/nature13763},
}

@article{AlbrechtPRB2015,
    author  = {Albrecht, Florian and Fleischmann, Martin and Scheer, Manfred and Gross, Leo and Repp, Jascha},
    title   = {Local tunneling decay length and {K}elvin probe force spectroscopy},
    journal = {Physical Review B},
    year    = {2015},
    volume  = {92},
    number  = {23},
    pages   = {235443},
    doi     = {10.1103/PhysRevB.92.235443}
}

@article{AugustinPRB2000,
    author  = {Augustin, J. and Eyert, V. and B{\"o}ker, Th. and Frentrup, W. and Dwelk, H. and Janowitz, C. and Manzke, R.},
    title = {Electronic band structure of the layered compound {T}$_{d}$-{WTe$_2$}},
    journal = {Physical Review B},
    year    = {2000},
    volume  = {62},
    number  = {16},
    pages   = {10812},
    doi     = {10.1103/PhysRevB.62.10812}
}

@article{BrownActaCryst1966,
	author = {Brown, B. E.},
	title = {The crystal structures of {WTe2} and high-temperature {MoTe2}},
	journal = {Acta Crystallographica},
	year = {1966},
	volume = {20},
	number = {2},
	pages = {268--274},
	doi = {10.1107/S0365110X66000513},
}

@article{BrunoPRB2016,
	author = {Bruno, F. Y. and Tamai, A. and Wu, Q. S. and Cucchi, I. and Barreteau, C. and de la Torre, A. and McKeown Walker, S. and Riccò, S. and Wang, Z. and Kim, T. K. and Hoesch, M. and Shi, M. and Plumb, N. C. and Giannini, E. and Soluyanov, A. A. and Baumberger, F.},
	title = {Observation of large topologically trivial Fermi arcs in the candidate type-{II} Weyl semimetal {WTe$_2$}},
    journal = {Physical Review B},
	year = {2016},
	volume = {94},
	number = {12},
	pages = {121112},
	doi = {10.1103/PhysRevB.94.121112},
}

@article{DaiPRB2015,
    author  = {Dai, Y. M. and Bowlan, J. and Li, H. and Miao, H. and Wu, S. F. and Kong, W. D. and Shi, Y. G. and Trugman, S. A. and Zhu, J.-X. and Ding, H. and Taylor, A. J. and Yarotski, D. A. and Prasankumar, R. P.},
    title   = {Ultrafast carrier dynamics in the large-magnetoresistance material {WTe$_2$}},
    journal = {Physical Review B},
    year    = {2015},
    volume  = {92},
    number  = {16},
    pages   = {161104},
    doi     = {10.1103/PhysRevB.92.161104}
}

@article{DruekePRB2021,
    author  = {Drueke, Elizabeth and Yang, Junjie and Zhao, Liuyan},
    title = {Observation of strong and anisotropic nonlinear optical effects through polarization-resolved optical spectroscopy in the type-{II} {W}eyl semimetal {T}$_{d}$-{WTe$_2$}},
    journal = {Physical Review B},
    year    = {2021},
    volume  = {104},
    number  = {6},
    pages   = {064304},
    doi     = {10.1103/PhysRevB.104.064304}
}

@article{FeiNat2018,
    author  = {Fei, Z. and Zhao, W. and Palomaki, T. A. and Sun, B. and Miller, M. K. and Zhao, Z. and Yan, J. and Xu, X. and Cobden, D. H.},
    title   = {Ferroelectric switching of a two-dimensional metal},
    journal = {Nature},
    year    = {2018},
    volume  = {560},
    number  = {7718},
    pages   = {336},
    doi     = {10.1038/s41586-018-0336-3}
}

@article{hoffmanScience2002,
	author = {Hoffman, J. E. and McElroy, K. and Lee, D.-H. and Lang, K. M and Eisaki, H. and Uchida, S. and Davis, J. C.},
	doi = {10.1126/science.1072640},
	journal = {Science},
	pages = {1148},
	title = {Imaging {Quasiparticle} {Interference} in {Bi$_2$Sr$_2$CaCu$_2$O$_{8+d}$}},
	volume = {297},
	year = {2002},
}

@article{HytchUltramicroscopy1998,
    author  = {H{\"y}tch, M. J. and Snoeck, E. and Kilaas, R.},
    title   = {Quantitative measurement of displacement and strain fields from {HREM} micrographs},
    journal = {Ultramicroscopy},
    year    = {1998},
    volume  = {74},
    pages   = {131},
    doi     = {10.1016/S0304-3991(98)00035-7}
}

@article{JhaAIPA2018,
    author  = {Jha, R. and Onishi, S. and Higashinaka, R. and Matsuda, T. D. and Ribeiro, R. A. and Aoki, Y.},
    title   = {Anisotropy in the electronic transport properties of {W}eyl semimetal {WTe$_2$} single crystals},
    journal = {AIP Advances},
    year    = {2018},
    volume  = {8},
    number  = {10},
    pages   = {101332},
    doi     = {10.1063/1.5043063}
}

@article{KawaharaAPEX2017,
    author  = {Kawahara, Kazuaki and Ni, Zeyuan and Arafune, Ryuichi and Shirasawa, Tetsuroh and Lin, Chun-Liang and Minamitani, Emi and Watanabe, Satoshi and Kawai, Maki and Takagi, Noriaki},
    title   = {Surface structure of novel semimetal {WTe$_2$}},
    journal = {Applied Physics Express},
    year    = {2017},
    volume  = {10},
    number  = {4},
    pages   = {045702},
    doi     = {10.7567/APEX.10.045702}
}

@article{KimPRB2017,
    author  = {Kim, Hyun-Jung and Kang, Seoung-Hun and Hamada, Ikutaro and Son, Young-Woo},
    title   = {Origins of the structural phase transitions in {MoTe$_2$} and {WTe$_2$}},
    journal = {Physical Review B},
    year    = {2017},
    volume  = {95},
    number  = {18},
    pages   = {180101},
    doi     = {10.1103/PhysRevB.95.180101}
}

@article{KwonApplSurfSci2020,
    author  = {Kwon, H. and Ji, B. and Bae, D. and Lee, J.-H. and Park, H. J. and Kim, D. H. and Kim, Y.-M. and Son, Y.-W. and Yang, H. and Cho, S.},
    title   = {Role of anionic vacancy for active hydrogen evolution in {WTe$_2$}},
    journal = {Applied Surface Science},
    year    = {2020},
    volume  = {515},
    pages   = {145972},
    doi     = {10.1016/j.apsusc.2020.145972}
}

@article{LawlerNature2010,
    author  = {Lawler, M. J. and Fujita, K. and Lee, J. and Schmidt, A. R. and Kohsaka, Y. and Kim, C. K. and Eisaki, H. and Uchida, S. and Davis, J. C. and Sethna, J. P. and Kim, E.-A.},
    title   = {Intra-unit-cell electronic nematicity of the high-{T}$_\mathrm{c}$ copper-oxide pseudogap states},
    journal = {Nature},
    year    = {2010},
    volume  = {466},
    pages   = {347},
    doi     = {10.1038/nature09169}
}

@article{LeeSciRep2015,
    author  = {Lee, C.-H. and Silva, E. C. and Calderin, L. and Nguyen, M. A. T. and Hollander, M. J. and Bersch, B. and Mallouk, T. E. and Robinson, J. A.},
    title   = {{T}ungsten {D}itelluride: a layered semimetal},
    journal = {Scientific Reports},
    year    = {2015},
    volume  = {5},
    number  = {1},
    pages   = {10013},
    doi     = {10.1038/srep10013}
}

@article{LiNatCommun2017,
    author  = {Li, Peng and Wen, Yan and He, Xin and Zhang, Qiang and Xia, Chuan and Yu, Zhi-Ming and Yang, Shengyuan A. and Zhu, Zhiyong and Alshareef, Husam N. and Zhang, Xi-Xiang},
    title   = {Evidence for topological type-{II} {W}eyl semimetal {WTe$_2$}},
    journal = {Nature Communications},
    year    = {2017},
    volume  = {8},
    number  = {1},
    pages   = {2150},
    doi     = {10.1038/s41467-017-02237-1}
}

@article{LiPRB2016,
    author  = {Li, Q. and Yan, J. and Yang, B. and Zang, Y. and Zhang, J. and He, K. and Wu, M. and Zhao, Y. and Mandrus, D. and Wang, J. and Xue, Q. and Chi, L. and Singh, D. J. and Pan, M.},
    title   = {Interference evidence for {R}ashba-type spin splitting on a semimetallic {WTe$_2$} surface},
    journal = {Physical Review B},
    year    = {2016},
    volume  = {94},
    number  = {11},
    pages   = {115419},
    doi     = {10.1103/PhysRevB.94.115419}
}

@article{LinACSNano2017,
    author  = {Lin, Chun-Liang and Arafune, Ryuichi and Liu, Ro-Ya and Yoshimura, Masato and Feng, Baojie and Kawahara, Kazuaki and Ni, Zeyuan and Minamitani, Emi and Watanabe, Satoshi and Shi, Youguo and Kawai, Maki and Chiang, Tai-Chang and Matsuda, Iwao and Takagi, Noriaki},
    title   = {Visualizing Type-{II} {W}eyl Points in {T}ungsten {D}itelluride by Quasiparticle Interference},
    journal = {ACS Nano},
    year    = {2017},
    volume  = {11},
    number  = {11},
    pages   = {11459},
    doi     = {10.1021/acsnano.7b06179}
}

@article{LiuNano2019,
    author  = {Liu, X. and Yang, Y. and Hu, T. and Zhao, G. and Chen, C. and Ren, W.},
    title = {Vertical ferroelectric switching by in-plane sliding of two-dimensional bilayer {WTe$_2$}},
    journal = {Nanoscale},
    year    = {2019},
    volume  = {11},
    number  = {40},
    pages   = {18575},
    doi     = {10.1039/C9NR05404A}
}

@article{LvPRL2017,
    author  = {Lv, Yang-Yang and Li, Xiao and Zhang, Bin-Bin and Deng, W. Y. and Yao, Shu-Hua and Chen, Y. B. and Zhou, Jian and Zhang, Shan-Tao and Lu, Ming-Hui and Zhang, Lei and Tian, Mingliang and Sheng, L. and Chen, Yan-Feng},
    title   = {Experimental Observation of Anisotropic {A}dler-{B}ell-{J}ackiw Anomaly in Type-{II} {W}eyl Semimetal {WTe$_{1.98}$} Crystals at the Quasiclassical Regime},
    journal = {Physical Review Letters},
    year    = {2017},
    volume  = {118},
    number  = {9},
    pages   = {096603},
    doi     = {10.1103/PhysRevLett.118.096603}
}

@article{mcelroyNature2003,
	author = {McElroy, K. and Simmonds, R. W. and Hoffman, J. E. and Lee, D.-H. and Orenstein, J. and Eisaki, H. and Uchida, S. and Davis, J. C.},
	title = {Relating atomic-scale electronic phenomena to wave-like quasiparticle states in superconducting {Bi$_2$Sr$_2$CaCu$_2$O$_{8+d}$}},
	journal = {Nature},
	year = {2003},
	volume = {422},
	pages = {592},
	doi = {10.1038/nature01496},
}

@article{MichaelsonJAP1977,
    author  = {Michaelson, H. B.},
    title   = {The work function of the elements and its periodicity},
    journal = {Journal of Applied Physics},
    year    = {1977},
    volume  = {48},
    number  = {11},
    pages   = {4729},
    doi     = {10.1063/1.323539}
}

@article{PengNatCommun2017,
	author = {Peng, Lang and Yuan, Yuan and Li, Gang and Yang, Xing and Xian, Jing-Jing and Yi, Chang-Jiang and Shi, You-Guo and Fu, Ying-Shuang},
	title = {Observation of topological states residing at step edges of {WTe} 2},
	journal = {Nature Communications},
	year = {2017},
	volume = {8},
	number = {1},
	pages = {659},
	doi = {10.1038/s41467-017-00745-8},
}

@article{QianSci2014,
    author  = {Qian, X. and Liu, J. and Fu, L. and Li, J.},
    title   = {Quantum spin {H}all effect in two-dimensional transition metal dichalcogenides},
    journal = {Science},
    year    = {2014},
    volume  = {346},
    number  = {6215},
    pages   = {1344},
    doi     = {10.1126/science.1256815}
}

@article{RocchinoNatCommun2024,
    author  = {Rocchino, Lorenzo and Balduini, Federico and Schmid, Heinz and Molinari, Alan and Luisier, Mathieu and Süß, Vicky and Felser, Claudia and Gotsmann, Bernd and Zota, Cezar B.},
    title   = {Magnetoresistive-coupled transistor using the {W}eyl semimetal {NbP}},
    journal = {Nature Communications},
    year    = {2024},
    volume  = {15},
    number  = {1},
    pages   = {710},
    doi     = {10.1038/s41467-024-44961-5}
}

@article{SanchesPRB2025,
  author = {S\'anchez-Barquilla, Raquel and Vega, Francisco Mart\'{\i}n and Ruiz, Alberto M. and Jo, Na Hyun and Herrera, Edwin and Baldov\'{\i}, Jos\'e J. and Ochi, Masayuki and Arita, Ryotaro and Bud'ko, Sergey L. and Canfield, Paul C. and Guillam\'on, Isabel and Suderow, Hermann},
   title = {Electronic band structure from quasiparticle interference and Landau quantization in WTe$_2$},
  journal = {Phys. Rev. B},
  year = {2025},  
  volume = {112},
  number = {16},
  pages = {165401},
  doi = {10.1103/7xnv-8wvt},
}

@article{SharmaSciAdv2019,
    author  = {Sharma, P. and Xiang, F.-X. and Shao, D.-F. and Zhang, D. and Tsymbal, E. Y. and Hamilton, A. R. and Seidel, J.},
    title   = {A room-temperature ferroelectric semimetal},
    journal = {Science Advances},
    year    = {2019},
    volume  = {5},
    number  = {7},
    pages   = {eaax5080},
    doi     = {10.1126/sciadv.aax5080}
}

@article{SieNat2019,
    author  = {Sie, E. J. and Nyby, C. M. and Pemmaraju, C. D. and Park, S. J. and Shen, X. and Yang, J. and Hoffmann, M. C. and Ofori-Okai, B. K. and Li, R. and Reid, A. H. and Weathersby, S. and Mannebach, E. and Finney, N. and Rhodes, D. and Chenet, D. and Antony, A. and Balicas, L. and Hone, J. and Devereaux, T. P. and Heinz, T. F. and Wang, X. and Lindenberg, A. M.},
    title   = {An ultrafast symmetry switch in a {W}eyl semimetal},
    journal = {Nature},
    year    = {2019},
    volume  = {565},
    number  = {7737},
    pages   = {61},
    doi     = {10.1038/s41586-018-0809-4}
}

@article{SoluyanovNat2015,
    author  = {Soluyanov, A. A. and Gresch, D. and Wang, Z. and Wu, Q. S. and Troyer, M. and Dai, X. and Bernevig, B. A.},
    title   = {Type-{II} {W}eyl semimetals},
    journal = {Nature},
    year    = {2015},
    volume  = {527},
    number  = {7579},
    pages   = {495},
    doi     = {10.1038/nature15768}
}

@article{SoranzioPRR2019,
    author  = {Soranzio, D. and Peressi, M. and Cava, R. J. and Parmigiani, F. and Cilento, F.},
    title   = {Ultrafast broadband optical spectroscopy for quantifying subpicometric coherent atomic displacements in {WTe$_2$}},
    journal = {Physical Review Research},
    year    = {2019},
    volume  = {1},
    number  = {3},
    pages   = {032033},
    doi     = {10.1103/PhysRevResearch.1.032033}
}

@article{SoranzioNPJ2D2022,
    author  = {Soranzio, D. and Savoini, M. and Beaud, P. and Cilento, F. and Boie, L. and D\"ossegger, J. and Ovuka, V. and Houver, S. and Sander, M. and Zerdane, S. and Abreu, E. and Deng, Y. and Mankowsky, R. and Lemke, H. T. and Parmigiani, F. and Peressi, M. and Johnson, S. L.},
    title   = {Strong modulation of carrier effective mass in {WTe$_2$} via coherent lattice manipulation},
    journal = {npj 2D Materials and Applications},
    year    = {2022},
    volume  = {6},
    number  = {1},
    pages   = {1},
    doi     = {10.1038/s41699-022-00347-z}
}

@article{TangNatPhys2017,
    author  = {Tang, S. and Zhang, C. and Wong, D. and Pedramrazi, Z. and Tsai, H.-Z. and Jia, C. and Moritz, B. and Claassen, M. and Ryu, H. and Kahn, S. and Jiang, J. and Yan, H. and Hashimoto, M. and Lu, D. and Moore, R. G. and Hwang, C.-C. and Hwang, C. and Hussain, Z. and Chen, Y. and Ugeda, M. M. and Liu, Z. and Xie, X. and Devereaux, T. P. and Crommie, M. F. and Mo, S.-K. and Shen, Z.-X.},
    title = {Quantum spin {H}all state in monolayer {1T}$'$-{WTe$_2$}},
    journal = {Nature Physics},
    year    = {2017},
    volume  = {13},
    number  = {7},
    pages   = {683},
    doi     = {10.1038/nphys4174}
}

@article{YangJPCL2018,
    author  = {Yang, Q. and Wu, M. and Li, J.},
    title   = {Origin of two-dimensional vertical ferroelectricity in {WTe$_2$} bilayer and multilayer},
    journal = {The Journal of Physical Chemistry Letters},
    year    = {2018},
    volume  = {9},
    number  = {24},
    pages   = {7160},
    doi     = {10.1021/acs.jpclett.8b03654}
}

@article{YuanPRB2018,
    author  = {Yuan, Y. and Yang, X. and Peng, L. and Wang, Z.-J. and Li, J. and Yi, C.-J. and Xian, J.-J. and Shi, Y.-G. and Fu, Y.-S.},
    title   = {Quasiparticle interference of {F}ermi arc states in the type-{II} {W}eyl semimetal candidate {WTe$_2$}},
    journal = {Physical Review B},
    year    = {2018},
    volume  = {97},
    number  = {16},
    pages   = {165435},
    doi     = {10.1103/PhysRevB.97.165435}
}

@article{ZhangNatRevMater2023,
    author  = {Zhang, Dawei and Schoenherr, Peggy and Sharma, Pankaj and Seidel, Jan},
    title   = {Ferroelectric order in van der {W}aals layered materials},
    journal = {Nature Reviews Materials},
    year    = {2023},
    volume  = {8},
    number  = {1},
    pages   = {25},
    doi     = {10.1038/s41578-022-00484-3}
}

@article{ZhangPRB2017,
    author  = {Zhang, W. and Wu, Q. and Zhang, L. and Cheong, S.-W. and Soluyanov, A. A. and Wu, W.},
    title   = {Quasiparticle interference of surface states in the type-{II} {W}eyl semimetal {WTe$_2$}},
    journal = {Physical Review B},
    year    = {2017},
    volume  = {96},
    number  = {16},
    pages   = {165125},
    doi     = {10.1103/PhysRevB.96.165125}
}

@article{ZhengPRL2016,
    author  = {Zheng, Hao and Bian, Guang and Chang, Guoqing and Lu, Hong and Xu, Su-Yang and Wang, Guangqiang and Chang, Tay-Rong and Zhang, Songtian and Belopolski, Ilya and Alidoust, Nasser and Sanchez, Daniel S. and Song, Fengqi and Jeng, Horng-Tay and Yao, Nan and Bansil, Arun and Jia, Shuang and Lin, Hsin and Hasan, M. Zahid},
    title   = {Atomic-Scale Visualization of Quasiparticle Interference on a type-{II} {W}eyl Semimetal Surface},
    journal = {Physical Review Letters},
    year    = {2016},
    volume  = {117},
    number  = {26},
    pages   = {266804},
    doi     = {10.1103/PhysRevLett.117.266804}
}

\end{document}